\documentclass[prl,aps,10pt,twocolumn,nopreprintnumbers,showpacs,tightenlines,floatfix,nofootinbib,amsfonts,amssymb]{revtex4-1}
\usepackage{epsfig}

\newcommand{\beqn}{\begin{eqnarray}}
\newcommand{\eeqn}{\end{eqnarray}}
\newcommand{\eq}[1]{(\ref{#1})}

\begin{document}

\title{Free magnetized knots of parity-violating deconfined matter in heavy-ion collisions}

\author{M. N. Chernodub}\thanks{On leave of absence from ITEP, Moscow, Russia.}
\affiliation{Laboratoire de Math\'ematiques et Physique Th\'eorique,
Universit\'e Fran\c{c}ois-Rabelais Tours,
F\'ed\'eration Denis Poisson - CNRS,
Parc de Grandmont, 37200 Tours, France\\
DMPA, University of Gent, Krijgslaan 281, S9, B-9000 Gent, Belgium}

\begin{abstract}
We show that the local parity violation in the quark-gluon plasma supports existence of free (meta)stable knots of deconfined
hot quark matter stabilized by superstrong magnetic fields. The magnetic field in the knots resembles the spheromak plasma state
of the magnetic confinement approach to nuclear fusion. The size of the knot is quantized, being inversely proportional
to the chiral conductivity of the quark-gluon plasma. The parity symmetry is broken inside the knot.
Particles produced in the decays of the knots have unusual azimuthal distribution and specific flavor content.
We argue that these knots may be created in noncentral heavy-ion collisions.
\end{abstract}


\date{February 7, 2010}

\maketitle

\section{Motivation}

Heavy-ion collisions may generate very strong magnetic fields with the magnitude of the order of the QCD scale.
The lowest bound of the maximal strength of the magnetic field $B$ that can emerge in noncentral collisions of Pb--Pb ions
at the Large Hadron Collider (LHC) at CERN was estimated in~\cite{ref:estimations} to be
\beqn
e B^{\mathrm{max}}_{\mathrm{LHC}} \gtrsim 15 \, m_\pi^2\,,
\label{eq:estimation}
\eeqn
where $e=|e|$ is the absolute value of the electron charge and $m_\pi \approx 140\, {\mathrm{MeV}}$ is the pion mass.
This superstrong long-ranged magnetic field acts as an external field on electrically charged quarks and antiquarks
that constitute the hot and dense quark-gluon matter created in the collision.

Theoretically, the magnetic field of the strength~\eq{eq:estimation} may lead to various nonperturbative effects such that
(i) a modification of the phase diagram of QCD via inequivalent shifts of both the chiral~\cite{ref:chiral} and deconfining finite-temperature
transitions~\cite{Agasian:2008tb}; (ii) an enhancement of the chiral condensate~\cite{ref:chiral:condensate} via an enforcement
of the dynamical chiral symmetry breaking~\cite{ref:magnetic:catalysis}; (iii) an anomaly-mediated creation of a stack of parallel domain
walls made of neutral pions in a dense plasma~\cite{Son:2007ny}; etc.

Strikingly, the superstrong hadron-scale magnetic field has already exposed
itself experimentally. Recently, the STAR collaboration has found
possible signatures of parity violation in locally macroscopic
domains of hot, strongly interacting matter created in
heavy-ion collisions at the RHIC
experimental facility~\cite{ref:RHIC:experiment}. Such domains may arise due to a nonvanishing expectation
value of the gluon topological charge that breaks both parity (P) and charge-parity (CP) symmetries.
In the hot quark-gluon plasma the spontaneous generation of the parity-violating domains -- suggested
theoretically in Ref.~\cite{Kharzeev:1998kz} -- may be revealed with the help of a very strong magnetic field
which is created in a noncentral heavy-ion collision as well. The key point is that in the parity-violating background the
external magnetic field leads to generation of an electric current. This unusual phenomenon -- used in the
experiment~\cite{ref:RHIC:experiment} to reveal the local parity violations via electric charge fluctuations --
is now known as the Chiral Magnetic Effect (CME)~\cite{ref:CME:2007,ref:CME:2008}. There exist also a numerical evidence
of existence of the CME based on the lattice simulations of QCD~\cite{Buividovich:2009wi}.

The essence of the CME may be formulated in the following way: if
a magnetic field $\vec B$ is applied to a system characterized by an asymmetry between the number of right-
and left-handed fermions (quarks), then an electric current $\vec j$ is induced along the magnetic
field axis~\cite{ref:CME:2008}:
\beqn
{\vec j} = \sigma_\chi {\vec B}\,.
\label{eq:CME}
\eeqn
The external electric field is assumed to be absent, $\vec E = 0$.

In \eq{eq:CME} the quantity $\sigma_\chi$ is the chiral conductivity~\cite{ref:CME:2008} and
\beqn
j_\mu = \sum_f e_f {\bar \psi}_f \gamma_\mu \psi_f\,,
\label{eq:J}
\eeqn
is the conserved electromagnetic four-current of the quarks. The vector $\vec j$
represents the spatial components of \eq{eq:J}.
In Eq.~\eq{eq:J} $e_q$ and $\psi_f$ are the electric charge
and the field of the $f^{\mathrm{th}}$ quark flavor, respectively, and $\gamma_\mu$ is the Dirac matrix.
The left-right quark asymmetry in the quark-gluon plasma -- that is needed for a realization of the CME~\eq{eq:CME} --
may be induced by various parity-odd (topological) effects~\cite{ref:CME:2008}.

The parity-odd law~\eq{eq:CME} can be contrasted with the ordinary
Ohm's law, ${\vec j} = \sigma {\vec E}$,
which is certainly a parity-even feature of the system. Here $\sigma$
is the conventional conductivity, and
the external magnetic field is assumed to be absent, $\vec B = 0$.

We show that the presence of the chiral conductivity~\eq{eq:CME} -- induced by topological QCD effects --
may lead to existence of (meta)stable knots made of the magnetic field that is
frozen in the hot quark-gluon plasma in a magnetohydrodynamic equilibrium.
In general, knotted configurations are of wide interest in many areas of physics.
One can encounter them in particle physics~\cite{Faddeev:Niemi:1997}, fluid dynamics~\cite{Moffat:1969},
plasma physics~\cite{Berger:1999}, condensed matter physics~\cite{Babaev:2001jt}, and classical field
theory~\cite{Radu:2008pp,Battye:Sutcliffe:1998}. Knotted light solutions to the vacuum Maxwell equations
are known theoretically~\cite{ref:knotted:light1}, and very recently they were discovered experimentally in
laser beams in a complementary approach of Ref.~\cite{ref:knotted:light2}.
Here we show that free knotted objets can be created in heavy-ion collisions as well.

\section{Knot solutions in parity-odd plasma}

The classical Maxwell equations of motion are
\beqn
\partial_\mu F^{\mu\nu} = j^\nu\,, \qquad \partial_\mu {\widetilde F}^{\mu\nu} = 0\,,
\label{eq:Maxwell}
\eeqn
where $F^{\mu\nu}$ is the electromagnetic field strength tensor,
${\widetilde F}^{\mu\nu} = (1/2) \varepsilon^{\mu\nu\alpha\beta} F_{\alpha\beta}$ is its dual,
and $j_\mu$ is the electric current density~\eq{eq:J}.
It is convenient to rewrite Eqs.~\eq{eq:Maxwell} in terms of electric $\vec E$ and magnetic $\vec B$ fields:
\beqn
\vec{\nabla} \times {\vec B} - \frac{\partial {\vec E}}{\partial t} = \vec j\,,
\label{eq:m1}\\
\vec{\nabla} \cdot {\vec E} = \rho\,,
\label{eq:m2}\\
\vec{\nabla} \times {\vec E} + \frac{\partial {\vec B}}{\partial t} = 0\,,
\label{eq:m3}\\
\vec{\nabla} \cdot {\vec B} = 0\,,
\label{eq:m4}
\eeqn
where $\rho = j_0$ is the electric charge density.

We are interested in static electrically neutral solutions, and therefore we set
\beqn
\vec E = 0\,, \qquad \rho = 0\,, \qquad \frac{\partial {\vec B}}{\partial t} = 0\,.
\label{eq:neutrality}
\eeqn
The quark-gluon plasma is conducting, and therefore the Maxwell equations should be supplemented with the laws,
that account for effects of the medium on the propagation of the electric current. In the absence of the bulk
electric field~\eq{eq:neutrality} this transport property is determined by the CME, Eq.~\eq{eq:CME}.

The requirement~\eq{eq:neutrality} is automatically consistent with Eqs.~\eq{eq:m2} and \eq{eq:m3},
while Eqs.~\eq{eq:m1} and \eq{eq:m4} give us a simpler system of equations:
\beqn
\vec{\nabla} \times {\vec B} & = & \sigma_\chi {\vec B}\,,
\label{eq:curl}\\
\vec{\nabla} \cdot {\vec B} & = & 0\,.
\label{eq:div}
\eeqn

We are looking for solutions of Eqs.~\eq{eq:curl} and \eq{eq:div} inside a ball of a certain radius $R$.
As we discuss below, the quark matter inside the ball, $r<R$ should be in a deconfined phase while the state outside the
ball corresponds to the confinement phase. Therefore, the electric current $\vec j$ in a stable configuration must
always be tangent to the boundary of the ball, $\vec j \cdot {\hat{r}} = 0$. Here
the unit vector in the $r$ direction, ${\hat{r}}$, represents the outward-normal vector to the boundary of the ball.
Then Eq.~\eq{eq:CME} gives us the following boundary condition:
\beqn
{\hat{r}} \cdot \vec B (\vec r) \Bigl{|}_{r=R} = 0\,.
\label{eq:boundary}
\eeqn

The system of equations \eq{eq:curl} and \eq{eq:div} is very well known because it plays an important role in usual plasma physics.
The corresponding solutions are usually called Chandrasekhar-Kendall states following the authors who stressed
importance of these solutions for astrophysical plasmas~\cite{CK:1957}. In the theory of nuclear fusion plasmas,
the eigenvectors of \eq{eq:curl} and \eq{eq:div} are known as Taylor states~\cite{Taylor:1974}. Such states are
ultimate minimum-energy configurations in the plasma equilibrium.

Mathematically, Eqs.~\eq{eq:curl}, \eq{eq:div} and \eq{eq:boundary} define an eigensystem
of the curl operator ``$\vec{\nabla} \times$''. Usually, an eigenvalue problem is
formulated in a fixed domain of space, so that the unknowns are eigenvalues. In our case
the eigenvalue in Eq.~\eq{eq:curl} is fixed by the known chiral conductivity $\sigma_\chi$,
which is assumed, for the sake of simplicity, to be coordinate-independent. The unknown
quantity is the radius of the ball, which enters the system of equations
via the boundary condition~\eq{eq:boundary}.

In the spherical domain the curl operator has a discrete set of quantized eigenvalues and eigenvectors,
that were described in details in Refs.~\cite{CK:1957,Spheromak:1979,Cantarella:2000}. Adopting these results to our
notations, we find that Eqs.~\eq{eq:curl}, \eq{eq:div}, \eq{eq:boundary} have solutions
in spherical balls of the quantized radii,
\beqn
R^{(l)}_k = \frac{\kappa^{(l)}_k}{\sigma_\chi}\,, \qquad l,k=1,2,3,\dots\,,
\eeqn
where $\kappa^{(l)}_k$ is the $k$th positive zero of the Bessel function $J_{n+1/2}(\kappa)$.
In our paper we are interested in the knot solution with the smallest possible radius
\beqn
R_0 = \frac{\kappa_0}{\sigma_\chi}\,,\qquad \kappa_0 \equiv  \kappa^{(1)}_1 = 4.49341\dots\,.
\label{eq:R0}
\eeqn
In the spherical coordinates the corresponding magnetic field configuration
is
\beqn
{\vec B}(r,\theta,\varphi) & = & B_0 \bigl[u(\sigma_\chi r,\theta) \, {\hat{r}} + v(\sigma_\chi r,\theta) \, {\hat{\theta}}
\nonumber\\
& & \hskip 23.5mm + w(\sigma_\chi r,\theta) \, {\hat{\varphi}}\bigr]\,,
\label{eq:B:solution}
\eeqn
for $r\leqslant R_0$, and ${\vec B}(r,\theta,\varphi) = 0$ for $r>R_0$. In Eq.~\eq{eq:B:solution}
\beqn
u(\xi,\theta) & = & \frac{2}{\xi^3} \left(\sin \xi - \xi \, \cos \xi\right) \cos \theta\,,
\label{eq:u}\\
v(\xi,\theta) & = & \frac{1}{\xi^3} \left(\sin \xi - \xi \, \cos \xi - \xi^2 \, \sin \xi \right) \sin \theta\,,
\label{eq:v}\\
w(\xi,\theta) & = & \frac{1}{\xi^2} \left(\sin \xi - \xi \, \cos \xi \right) \sin \theta\,,
\label{eq:w}
\eeqn
and ${\hat{\theta}}$ and ${\hat{\varphi}}$ are the unit vectors in the azimuthal and polar directions,
respectively. The azimuthal angle, $\theta \in (0,\pi)$ is inclination from the $z$ axis, while the polar angle,
$\varphi \in (-\pi,\pi)$, operates in the $xy$ plane. The coordinate-independent factor $B_0$ in Eq.~\eq{eq:B:solution}
is a free parameter.

\begin{figure}[ht]
\begin{center}
\vskip -4mm
  \includegraphics[width=80mm, angle=0]{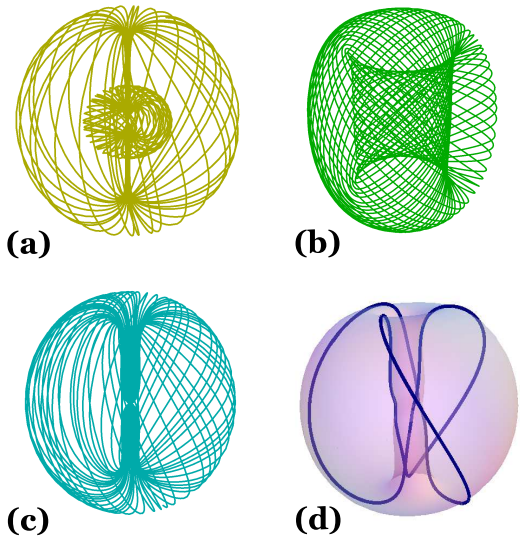}
\end{center}
\vskip -5mm
  \caption{Typical structures formed by the lines of the magnetic field and, equivalently, by the electric currents.
  The lines wind around different nested tori inside the same knot solution~\eq{eq:B:solution}--\eq{eq:w}.
  The figures are described in the text.}
  \label{fig:field:lines}
\end{figure}

The structure of the magnetic field of the solution \eq{eq:B:solution}--\eq{eq:w} is rather complicated.
The solution consist of an infinite number of nested (deformed) tori, and the degree of the deformation
varies with the torus size. The magnetic field lines wrap around each
torus and produce multiple self-linkages, what is a typical feature of all Chandrasekhar-Kendall
states~\cite{ref:knotted:light1}. Therefore we call the solution~\eq{eq:B:solution}--\eq{eq:w} as ``knot''.
The structure of typical magnetic lines and electric currents in the knot is illustrated in Fig.~\ref{fig:field:lines}.

It is convenient to label each structure by specifying the maximal size $r_{\mathrm{max}}$
of the (deformed) torus in units of the overall knot size $R_0$, Eq.~\eq{eq:R0}.
In the largest structures the field wind around two tori simultaneously
[Fig.~\ref{fig:field:lines}(a), $r_{\mathrm{max}}/R_0 = 0.96$].
As the size becomes smaller, the structure of the magnetic fields gets a conventional torus shape
[Fig.~\ref{fig:field:lines}(b), $r_{\mathrm{max}}/R_0 = 0.91$].
Yet another torus shape can be found at lower radii
[Fig.~\ref{fig:field:lines}(c), $r_{\mathrm{max}}/R_0 = 0.77$].
At particular radii the magnetic field lines form well-known knotted structures.
For example, at $r_{\mathrm{max}}/R_0 = 0.88566$, the magnetic lines are trefoil knots,
which wrap around a torus, Fig.~\ref{fig:field:lines}(d).

Notice, that despite the knot occupies an ideally spherical region, its interior is
only axially (not spherically) symmetric. Moreover, the knot is a parity-odd
object: its reflected image cannot be superimposed on itself. A mirror image of a
left-handed knot is a right-handed knot.

In an astrophysical applications the knot \eq{eq:B:solution}-\eq{eq:w} was used to model magnetic field
patterns of the Crab Nebula~\cite{Woltjer:1958}. At much higher energies, in the nuclear fusion concept,
the solution \eq{eq:B:solution}-\eq{eq:w} is considered as a relaxed minimum energy state of a plasma in a spheromak device
(i.e., inside a spherical shell that confines the plasma)~\cite{Spheromak:1979,Review:Taylor}.  In general, the solutions of the system of
equations \eq{eq:curl}, \eq{eq:div} and \eq{eq:boundary} are the ``natural end configurations'' in a magnetically dominated plasma,
because the magnetic turbulence drives such plasma towards the spheromak state~\cite{Review:Taylor,Woltjer:1958}. We suggest that similar stabilized
configurations may arise in the quark-gluon plasma created in heavy-ion collisions.

\section{Basic physical features of knots}

{\it Phase.} The interior of the knot should be in a deconfined state with unbroken
chiral symmetry. The deconfinement is needed for the quarks to flow freely inside the knot,
so that they should be able to produce stabilizing magnetic fields. The chirally restored symmetry is also very important
because in this state the chiral condensate is zero. A presence of the chiral condensate would wash out the imbalance
between the left-handed and the right-handed quarks, leading to vanishing chiral
conductivity~\cite{ref:CME:2008}, $\sigma_\chi=0$. The very existence of the knot is guaranteed by a nonvanishing value of $\sigma_\chi$,
and therefore the knot cannot exist if its interior is in the chirally broken phase. The chirally restored and simultaneously
deconfined phase is expected to be produced at the RHIC facility and at the ALICE/LHC experiment~\cite{ref:ALICE},
so that the knots have a chance to be created there.

\begin{figure}[ht]
\begin{center}
\vskip -4mm
  \includegraphics[width=70mm, angle=0]{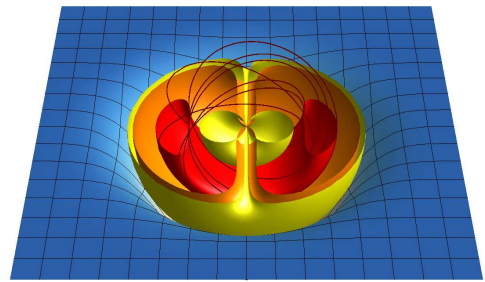}
\end{center}
\vskip -20mm
  \caption{A free hot knotted solution~\eq{eq:B:solution}--\eq{eq:w} in the cold QCD vacuum.
  The tori that host the magnetic lines of Figs.~\ref{fig:field:lines}(a)-(c), and some
  of the magnetic lines are shown.}
  \label{fig:qualitative}
\end{figure}

Generally, one may expect that the cold exterior outside the knot should provide an additional factor that stabilizes the knot.
The nonzero tension of the interface between the phases would squeeze the knot increasing the energy density of the magnetic field,
and generating a backreaction force from the inside. A qualitative illustration of the knot structure is presented in
Fig.~\ref{fig:qualitative}.

\vskip 3mm
{\it The electric charge} of the knot is zero. Other important
parameters of the knot may be expressed via two free parameters of the solution~\eq{eq:B:solution}--\eq{eq:w}:
the magnetic field scale $B_0$ and the chiral conductivity~$\sigma_\chi$.

\vskip 3mm
{\it The knot radius} $R_0$ is given by Eq.~\eq{eq:R0}.

\vskip 3mm
{\it The total magnetic energy} of the knot is
\beqn
E_{\mathrm{magn}} = \frac{1}{2} \int {\mathrm d}^3 r \, {\vec B}^2(\vec r) = C_E \frac{B_0^2}{\sigma_\chi^3}
\equiv \frac{C_E}{\kappa_0^3} B_0^2 R_0^3\,, \quad
\label{eq:Energy}
\eeqn
where $C_E = (4 \pi/3) \kappa_0 \sin^2 \kappa_0 = 17.9337 \dots\,$.

\vskip 3mm
{\it The total magnetic moment} of the knot is
\beqn
{\vec M} = \int {\mathrm d}^3 r \, \frac{1}{2} {\vec r} \times {\vec j} = C_M \frac{B_0}{\sigma_\chi^3} {\hat z}
\equiv \frac{C_M}{\kappa_0^3} B_0 R_0^3 \cdot {\hat z}\,, \quad
\label{eq:MagneticMoment}
\eeqn
where $C_M = \pi^2 [1-(1+\kappa_0^2/2) \cos \kappa_0] = 33.6592 \dots\,$ and
$\hat z$ is the unit vector along the $z$-axis.

\vskip 3mm
{\it The flavor content} of the knot is constrained by the requirement of
stability of the knot with respect to direct quark-antiquark annihilations.
This implies that the interior of the knot may consist of either $u$, $d$ and $s$
quarks, {\it or} $\bar u$, $\bar d$ and $\bar s$ antiquarks. A mixed interior, say,
$u$, $d$ and $\bar s$ is less favorable energetically. The presence of
heavier quarks -- with the masses heavier than a few $T_c$ --
is dynamically suppressed.

The condition of the local electric neutrality
[here $n_f = n_f(x)$ is the density of quarks of $f^{\mathrm{th}}$ flavor],
\beqn
\rho = 0\,,
\qquad
\rho \equiv \sum_f e_f n_f = \frac{2e}{3} n_u - \frac{e}{3} n_d - \frac{e}{3}  n_s\,,
\qquad
\eeqn
sets the constraint on the total quark numbers $N_f$:
\beqn
2 N_u = N_d + N_s
\quad \mbox{with} \quad
N_f = \int {\mathrm d}^3 r \, n_f(\vec r)\,.
\label{eq:n:flavor}
\eeqn
A similar relation holds for the antiquarks in a knot made of antimatter.
Due to relatively large mass of the strange quark we expect that
the strange fraction, $\beta_s = N_s/(N_u + N_d + N_s)$,
should be low, $\beta_s \ll 1$. Thus, the flavor composition of the knot's interior
should be similar to the one of a neutron star, $N_d \simeq 2 N_u$.

\vskip 3mm
{\it The minimal baryon charge} of the knot,
\beqn
Q_B = \frac{1}{3} \sum_f N_f = \frac{1}{3} (N_u + N_d + N_s)\,,
\label{eq:QB}
\eeqn
can be estimated as follows. The local electric current is
\beqn
{\vec j}(\vec r) & \equiv & \sum_f e_f n_f(\vec r) {\vec v}_f(\vec r)
\label{eq:j:flavor} \\
& = & \frac{2e}{3} n_u {\vec v}_u - \frac{e}{3} n_d {\vec v}_d - \frac{e}{3} n_s {\vec v}_s\,,
\nonumber
\eeqn
where ${\vec v}_f$ is the speed of the quarks of $f^{\mathrm{th}}$ flavor. A minimal value of the baryonic charge is
reached in the case of the maximal ``charge-parity polarization'', i.e. when all positively charged quarks ($u$) are,
say, left-handed while the negatively-charged quarks ($d$ and $s$) are all right-handed. In this case the CME-driven~\eq{eq:CME}
current takes its maximum value at a fixed baryon charge. In the strong external magnetic field positively charged $u$ quarks and
negatively charged $d$ and $s$ quarks move in opposite directions. Ignoring a contribution of the strange quarks (i.e.,
setting $\beta_s=0$), one gets ${\vec v}_u = - {\vec v}_d$ for the light quarks which travel almost with the speed of light,
$|{\vec v}_{u,d}| = 1$. Using~\eq{eq:n:flavor}, we get the absolute value of the current:
\beqn
j(\vec r) \equiv |{\vec j}(\vec r)| = \frac{4 e}{3} n_u(\vec r) = \frac{2 e}{3} n_d (\vec r)\,.
\eeqn
Equation~\eq{eq:CME} then allows us to express the quark's densities $n_{u,d}$ via the magnetic field
of the solution~\eq{eq:B:solution}--\eq{eq:w}. Then the baryon charge of the knot~\eq{eq:QB} is:
\beqn
Q_B = \frac{3}{4 e} \int {\mathrm d}^3 r \, j(\vec r) = C_Q \frac{B_0}{e \sigma_\chi^2}
\equiv \frac{C_Q}{\kappa_0^2} \frac{B_0 R_0^2}{e}\,, \quad
\eeqn
where $C_Q = 80.4854 \dots$. The knots made of antimatter have a negative baryonic charge,
$Q_B = - |Q_B|$.

\vskip 3mm
{\it In a real physical environment} of heavy-ion collisions the approximate characteristics of the knot
can be estimated as follows. At high temperature $T$ the chiral conductivity is~\cite{ref:CME:2008}:
\beqn
\sigma_\chi = \frac{3 \overline{\, q^2}}{2 \pi^2} \frac{|n_5|}{T^2 + \mu^2/\pi^2} \,, \qquad
\overline{\, q^2} = \frac{1}{N_f} \sum_f e_f^2\,,
\label{eq:sigma:T}
\eeqn
where $N_f$ is the total number of participating quark flavors, $\overline{\, q^2}$ is the
average squared electric charge of these flavors,  $\mu$ is the quark chemical
potential, and $n_5 = c_5 \cdot \mbox{fm}^{-3}$ is the chiral density given by the difference between
the number of left- and right-handed quarks per unit volume. Here $c_5 = 2 N_f Q_w$, where $Q_w$
is the topological winding number (accumulated) per fm${}^3$.

For a qualitative estimation of the knot's parameters in Pb-Pb collisions at ALICE, we take $N_f = 2$
(so that $\overline{\, q^2} = 5 \, e^2/18$) and we assume that scale of the finite chemical
potential is the same as the temperature scale, $\mu/\pi \sim T$. Thus, slightly above the
phase transition, $T \sim 200 \,\mbox{MeV}$, one gets $\sigma_\chi \sim 0.4 \, c_5 \, \mbox{MeV}$. Then
the knot radius~\eq{eq:R0} is $R_0 \sim (2\cdot 10^3/c_5)\,\mbox{fm}$. Typically~\cite{ref:CME:2007,ref:CME:2008},
$c_5 = (1-10)$ so that the typical radius of the knot is as small as $R \sim 200$\,fm. This length is
large compared to a geometrical scale ($\sim$10\,fm) of a typical heavy-ion collision, so that a creation
of such large knot is unlikely. However,
very large fluctuations of the topological charge in particular events are not excluded. Thus, in our qualitative estimation
we set $Q_w \sim 50$ per fm${}^3$, that gives us $c_5 \sim 200$ and $\sigma_\chi = 80\,\mbox{MeV}$.
As for the scale of the magnetic field $B_0$, we take a fraction of the peak value~\eq{eq:estimation}, $e B_0 \sim m_\pi^2$.
Then we get the radius~\eq{eq:R0}, the total magnetic energy~\eq{eq:Energy}, the total magnetic moment~\eq{eq:MagneticMoment}
in terms of the nuclear magneton $\mu_N = e/(2 m_p)$,
and the minimal baryon charge~\eq{eq:QB} of the knot, respectively:
\beqn
\begin{array}{rclcrcl}
R_0 & \sim & 10\,\mathrm{fm}\,,
& \quad &
E_{\mathrm{magn}} & \sim & 150\, \mathrm{GeV}\,, \\
|{\vec M}| & \sim & 2.5\cdot 10^4\, \mu_N\,,
& \quad &
Q_B & \sim & 2.5 \cdot 10^3\,.
\end{array}
\label{eq:phys}
\eeqn

The radius of the knot $R_0$ is of the order of the size of the lead nucleus. The magnetic energy of the knot $E_{\mathrm{magn}}$
is also a natural quantity being a bit smaller than the rest energy of the Pb nucleus. The corresponding energy density,
$\varepsilon_{\mathrm{magn}} = E_{\mathrm{magn}}/(4 \pi R_0^3/3) \sim 35 \mathrm{MeV}/\mathrm{fm}^3$ is much smaller
than the non-equilibrium energy density regime of ALICE, $\varepsilon \simeq (1-1000)\, \mathrm{MeV}/\mathrm{fm}^3$~\cite{ref:ALICE}.
The magnetic moment of the knot is rather large in nuclear units, while $|{\vec M}| \sim 15 \, \mu_B$
in terms of the Bohr magneton~$\mu_B$.

The baryon charge $Q_B$ of the knot~\eq{eq:phys} is six time larger than the total number
($\sim$400) of nuclei in the colliding lead ions. However, the total multiplicity of produced particles in Pb-Pb collisions is estimated
to be in the range of thousands~\cite{ref:ALICE}, so that the large baryon charge $Q_B$ of the knot may be within experimental reach.

The lifetime of the knot is difficult to estimate from general considerations since a proper analysis should
involve thermal magnetodynamic calculations in a so-far-unexplored parity-odd backgrounds. We hope that
this analysis can be done using a numerical approach.

\section{Geometry of the knot decay}

The knot may be observed via the particles produced by its decay. Besides the specific flavor content (the  $u$ and $d$ quarks
should appear in proportions 1:2), the flow of produced particles is rather unusual as we will see below.

\begin{figure}[ht]
\begin{center}
  \includegraphics[width=70mm, angle=0]{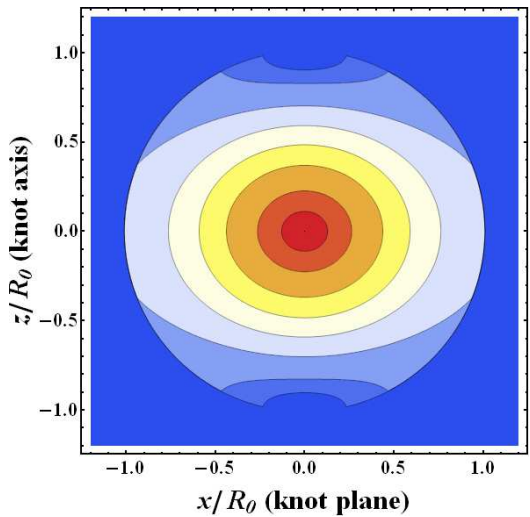}
\end{center}
  \caption{The contour plot of the strength of the electric currents $\varrho = |\vec j(\vec r)|/(\sigma_\chi B_0)$
  in the transverse $xz$ plane (determined by $y=0$) of the knot. The contour lines (from outer to inner)
  correspond to $\varrho=0.05,0.1,0.2,0.3,0.4,0.5,0.6,0.65$.}
  \label{fig:energy}
\end{figure}
Although the knot occupies the spherical region, the knot's interior has a lower (axial) symmetry.
The spatial distribution of the magnetic energy density has a form similar to the one of an oblate (disk-shaped) spheroid.
The short axis of the spheroid is parallel to the magnetic moment~\eq{eq:MagneticMoment} of the knot
($z$-axis in our notations), while the short axes are located in the plane of the knot ($xy$-plane).
A contour plot of the strength of the electric currents in a transverse $xz$ plane of the knot is shown in Fig.~\ref{fig:energy}.
The standard eccentricity of the electric currents in the knot is negative:
\beqn
\epsilon_j = \frac{\langle z^2 - x^2\rangle_j}{\langle z^2 + x^2\rangle_j} = -0.168 \dots\,,
\eeqn
where $\langle \dots \rangle_j = \int {\mathrm d} x {\mathrm d} z \dots \, j(x,z)$. In the contrary,
an initial geometry of a noncentral heavy-ion collision has an almond shape characterized by
a positive eccentricity.

A decay of a knot should produce particles that dominantly follow the current distribution of
the quark's current in the equilibrium state of the knot. Using the explicit knot solution~\eq{eq:B:solution}--\eq{eq:w}
one can calculate the distribution of the decay products
\beqn
\frac{{\mathrm d^2} N}{{\mathrm d} \phi \, {\mathrm d} \varphi} = \int {\mathrm{d}}^3 r \, |j(\vec r)|
\delta\Bigl({\hat n}(\phi,\varphi) - {\hat j}(\vec r)\Bigr)\,,
\label{eq:N1}
\eeqn
where ${\hat j}(\vec r) = {\vec j}(\vec r)/|{\vec j}(\vec r)|$ is the unit vector in the direction of the quark current
at point $\vec r$, and  ${\hat n}(\phi,\varphi)$ is the unit vector pointing in the direction determined by the polar angle
$\varphi$ and the new azimuthal angle $\phi = \pi/2 - \theta$. The angle $\phi$ gives
inclination from the $xy$ plane (we follow the standard notations used in physics of the heavy-ion collisions).

\begin{figure}[ht]
\begin{center}
  \includegraphics[width=80mm, angle=0]{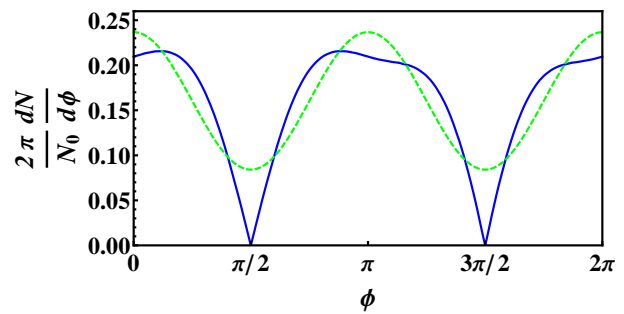}
\end{center}
\vskip -5mm
  \caption{Normalized azimuthal distribution of particles created in the knot decay (the solid blue line),
  and its elliptic component (the dashed green line).}
  \label{fig:azimuthal}
\end{figure}

An explicit calculation reveals that the distribution~\eq{eq:N1} is independent of the polar angle $\varphi$, so that
the decay of the knot is axially symmetric with respect to rotations around the $z$-axis.
However, the azimuthal distribution of the particles emitted from the decaying knot
is quite nontrivial (Fig.~\ref{fig:azimuthal}) as it differs significantly from a typical azimuthal distribution of
particles produced in heavy-ion collisions.

The azimuthal particle distribution can be expanded into a Fourier series,
\beqn
\frac{2\pi}{N_0 } \frac{{\mathrm d} N}{{\mathrm d} \phi} = 1 +
2 \sum_{k=1}^\infty \bigl[v_k \cos (k \phi) + w_k \sin (k \phi) \bigl]\,,
\label{eq:N:Fourier}
\eeqn
were $N_0$ is the total number of the particles produced in the knot decay. Symmetries of the knot imply
that $v_{2k-1}=w_{2k}=0$ with $k\in \mathbb{N}$.

The first five nonzero $v$- and $w$-coefficients of the Fourier expansion~\eq{eq:N:Fourier} are summarized
in Table~\ref{tbl:parameters}.
\begin{table}[htb]
\begin{center}
\caption{The Fourier flow coefficients of the particles produced in the knot decay.} \label{tbl:parameters}
\begin{tabular}{cccccc}
\hline
\hline
$n$ & 1 & 2 & 3 & 4 & 5 \\
\hline
$v_{2n}$    & 0.2378 & -0.1138 & 0.04425 & -0.02389 & 0.01505 \\
$w_{2n-1}$  & 0.001695  &  0.01619  & 0.005640 & -0.002061 & 0.001731 \\
\hline
\hline
\end{tabular}
\end{center}
\end{table}
The direct flow of the knot decay is zero, $v_1=0$, while the elliptic flow is rather large, $v_2 \approx 1/4$. The elliptic harmonic,
$1 + 2 v_2 \cos(2\phi)$, is shown in Fig.~\ref{fig:azimuthal} by the dashed green line. Notice, that the higher-order $v$-harmonics
are not small. For example, the ratio of fourth and (squared) second cosine coefficients is negative and large
in its absolute value,
\beqn
v_4/v_2^2 = -2.01 \dots\,,
\eeqn
contrary to the case of the heavy-ion collisions where one expects $v_4/v_2^2 \approx 0.5$~\cite{Gombeaud:2009ye}.

In the contrast with the heavy-ion collisions, the sine-terms of the expansion~\eq{eq:N:Fourier} are nonzero.
This fact indicates that the knot decay is asymmetric with respect to the $z \to -z$ flips: the distribution of
the emitted particles, say, above the $xy$ plane does not coincide with the distribution that is observed below the
plane. The $z$-axis asymmetry
is consistent with the fact that the magnetic moment of the knot is nonzero.

The form of the azimuthal distribution, Fig.~\ref{fig:azimuthal}, signals that the largest flux of the decay
products is emitted a bit above the knot's plane. The maximal flow is concentrated around the cone with
the azimuthal angle
\beqn
\phi_{\mathrm{max}} = 0.343\dots \quad {\mathrm{or}} \quad \phi_{\mathrm{max}} \approx 20^o\,.
\eeqn
The flow of the decay particles in the direction of the knot axis ($\phi=0,\pi$) is zero.

\begin{figure}[ht]
\begin{center}
  \includegraphics[width=70mm, angle=0]{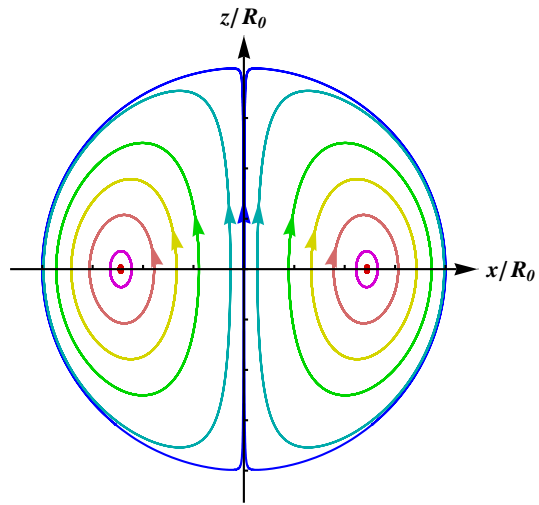}
\end{center}
\vskip -5mm
  \caption{The {\it projection} of the magnetic field lines on the $y=0$ cross-section of the knot.
  The lines have, in general, a $\hat y$ component (not shown).}
  \label{fig:cross}
\end{figure}

Finally, we notice that it is likely that the plane of the knot may tend to coincide with the reaction plane of a noncentral
heavy-ion collision that creates the knot. Indeed, the strongest magnetic field -- created in the center of the collision --
is perpendicular to the reaction plane while the strongest magnetic field of the knot
(observed in its center, Fig.~\ref{fig:energy}) is perpendicular to the knot's plane,
Fig.~\ref{fig:cross}. Since the knot is formed via a reconnection process of the lines of the magnetic field created in the
collision, the orientations of the magnetic field lines of the collision and of the knot are likely to be correlated.
This fact, in turn, implies a correlation of the corresponding planes.

\section{Summary}

In the summary, we outline our vision of the life of a typical magnetized knot.

\vskip 3mm
{\it Creation.} A noncentral collision of two heavy ions creates
a dense hot medium (plasma) of quarks and gluons pierced by lines of a strong magnetic field. The magnetic field
-- that is induced by moving charged ions and products of their collision -- is perpendicular to the reaction plane.
The dense quark-gluon matter expands and cools down, while (some of) the magnetic lines reconnect.
Simultaneously, fluctuations of the topological charge in the plasma lead to emergence of the local parity-violating
domains characterized by the imbalance between the left- and right-handed quarks. The violation of the parity symmetry in the background of
the magnetic field leads to the chiral magnetic effect, and the electric current start to flow along the lines of the magnetic
field.

\vskip 3mm
{\it Stabilization.} If the reconnected magnetic lines happen to be knotted, then the electric current creates a self-supporting
state, since the current flowing at the one segment of the knot supports the magnetic field and, consequently, the electric current
at other segments. Magnetic turbulence drives the plasma towards a minimum energy configuration that resembles the Taylor spheromak
state in the magnetically confined plasmas of nuclear fusion devices. Both the closeness of the knotted magnetic lines, the conservation
of the electric current,
and the squeezing effect of the knot boundary contribute to the stability of the knot. The magnetic lines hold the hot quark plasma
confined inside the knot. The size of the knot is expected to be of the order of some tens of fermi. Generally, the knot is an electrically
neutral system with a neutron-star-like flavor content: one $u$ quark comes with two $d$ quarks with a possible small admixture of $s$ quarks.

\vskip 3mm
{\it Decay.} As the knot cools down, the emerging chiral condensate starts to wash out the imbalance between the left- and right-handed
fermions and eventually leads to the decay of the knot. We cannot estimate the lifetime of the knot using general arguments, but we expect
that it should be much larger than a typical lifetime of a quark-gluon fireball created in the collision.
Primary products of the knot decay are $u$ and $d$ quarks coming out with the specific
ratio $1:2$. The emitted particles have the unusual azimuthal distribution characterized by a significant elliptic
flow $v_2 \approx  1/4$, relatively large (and sometimes negative) higher harmonics (i.e., $v_4/v_2^2 \approx - 2$),
and axially-symmetric broad cone-shaped distribution of the emitted particles. The aperture of the emission cone is
$180^o - 2\phi_{\mathrm{max}} \approx 140^o$. The specific flavor content and
azimuthal distribution of the decay products may serve as good experimental signatures of the knots at RHIC and ALICE/LHC experiments.

\vskip 3mm

\acknowledgments
The author is grateful to P.~Forgacs and A.J.~Niemi for useful discussions.

\end{document}